\documentclass[published]{JHEP3} 

\JHEP{00(2006)000}

\JHEPspecialurl{http://jhep.sissa.it/JOURNAL/JHEP3.tar.gz}

\usepackage{epsfig,multicol,bbm}

\newcommand\fverb{\setbox\pippobox=\hbox\bgroup\verb}
\newcommand\fverbdo{\egroup\medskip\noindent%
            \fbox{\unhbox\pippobox}\ }
\newcommand\fverbit{\egroup\item[\fbox{\unhbox\pippobox}]}

\def\dd{\displaystyle}

\newbox\pippobox

\title{ The Dirichlet Casimir effect for $\phi^4$ theory in (3+1) dimensions: A new renormalization approach}

\author{Reza Moazzemi, Maryam Namdar,
        and Siamak S. Gousheh\\
    Department of Physics, Shahid Beheshti University, Evin, Tehran
19839, Iran\\
    E-mail: \email{R-Moazzemi@sbu.ac.ir}, \email{m-namdar@sbu.ac.ir}, \email{ss-gousheh@sbu.ac.ir}}

\received{}        
\revised{} \accepted{}

\abstract{We calculate the next to the leading order Casimir effect
for a real scalar field, within $\phi^4$ theory, confined between
two parallel plates in three spatial dimensions with the Dirichlet
boundary condition. In this paper we introduce a systematic
perturbation expansion in which the counterterms automatically turn
out to be  consistent with the boundary conditions. This will
inevitably lead to nontrivial position dependence for physical
quantities, as a manifestation of the breaking of the translational
invariance. This is in contrast to the usual usage of the
counterterms in problems with nontrivial boundary conditions, which
are either completely derived from the free cases or at most
supplemented with the addition of counterterms only at the
boundaries. Our results for the massive and massless cases are
different from those reported elsewhere. Secondly, and probably less
importantly, we use a supplementary renormalization procedure, which
makes the usage of any analytic continuation techniques unnecessary.
}


\begin{document}

\section{Introduction}
During the last fifty years many papers have been written on the
Casimir effect. In this paper we introduce a new approach in regards
to the renormalization program. We use our approach in one of the
simplest nontrivial example possible, i.e. a real scalar field
confined between two parallel plates in 3+1 dimensions, with
$\phi^4$ self-interaction. As we shall see, our results for the next
to leading order term (NLO) differs significantly from what exists
in the literature. It is therefore suitable to start at the
beginning. In 1948 H.B.G. Casimir found a simple yet profound
explanation for the retarded van der Waals interaction
\cite{Casimir}. After a short time, he and D. Polder related this
effect to the change in the zero point energy of the quantum fields
due to the presence of nontrivial boundary conditions \cite{Polder}.
This energy has since been called the Casimir energy. The zero-order
energy in perturbation theory has been calculated for various fields
(see for example \cite{Wolfram,Svaiter}). Also the NLO correction,
which is usually called the first-order effect, has been computed
for various fields. For the electromagnetic field this correction is
said to be due to the following Feynman diagram
\raisebox{-2mm}{\includegraphics{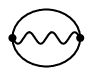}} , and has been computed first
by Bordag and collaborators \cite{brw}. However, note that this
correction is a two loop correction in this case and is ${\cal
O}(e^2)$. Moreover the two-loop radiative corrections for some
effective field theories have been investigated in \cite{rt,kr,km}.
Next, in the case of a real massive scalar field NLO correction to
the energy has been computed in
\cite{ford,kay,toms,lang,albu,Mostepanenko,Baron1,Baron2}. This
correction is a two loop correction in this case but is $\cal
O(\lambda)$. Moreover, N. Graham \emph{et al} used new approaches to
this problem by utilizing the phase shift of the scattering states
\cite{Graham}, or replacing the boundary conditions by an
appropriate potential term \cite{Graham2}. However, all of the
authors who use counterterms to calculate NLO correction to the
Casimir energy use the free counterterms (in the space between the
plates) by which we mean the ones that are relevant to the free
cases with no nontrivial boundary conditions, and are obviously
position independent. Only in \cite{albu} the author notes that in
certain cases, counterterms can depend on the distance between the
plates. The first use of nontrivial boundary conditions for the
renormalization programs in problems of this sort seems to be due to
Fosco and Svaiter \cite{Fosco}. These authors use free counterterms
in the space between the plates and place additional surface
counterterms at the boundaries. Later on various authors proposed
the use of exactly the same renormalization procedure for various
physical problems \cite{Nogu}. The first calculation for the NLO of
Casimir energy for the massive scalar field using this
renormalization program is done in ref.\cite{Caval}. We should note
that their results for the massless limit in 1+1 dimensions is
infinite similar to \cite{Baron1}, who used free counterterms only.
Moreover, the results for the massless limit in $2+1$ case reported
in ref.\cite{Baron1,Caval}, depend crucially on the order in which
the limits $d\to 2$ and $m\to 0$ are taken. When the order of the
limits are as shown, their result is infinite. When the order is
opposite, they get finite results which, surprisingly, contains the
Euler-Mascheroni number for both the massive and massless cases.
This is in contrast to the usual point of view that this number
should not appear in any results which reflect a physical quantity
\cite{Buras,Peskin}. We interpret these negative results as
indications that the use of free counterterms in these problems
might not be appropriate. It is also worth mentioning that all the
papers on the analogous calculations of the NLO corrections to the
mass of solitons, that we are aware of, use free counterterms (see
for example \cite{soliton,rebhan,dashen}). In references
\cite{rebhan} the authors used the mode number cutoff introduced by
R.F. Dashen (1974) \cite{dashen} to calculate the NLO Casimir
energy due to the presence of solitons.

In this paper, we present a systematic approach to the
renormalization program for problems which are amenable to
renormalized perturbation theory, and contain either nontrivial
boundary conditions or nontrivial (position dependent) backgrounds,
e.g. solitons, or both. Obviously all the n-point functions of the
theory will have in general nontrivial position dependence in the
coordinate representation. This is one of the manifestations of the
breaking of the translational symmetry. The procedure to deduce the
counterterms from the n-point functions in a renormalized
perturbation theory is standard and has been available for over half
a century. Using this, as we shall show, we will inevitably obtain
position dependent counterterms. Therefore, the radiative
corrections to all the input parameters of the theory, including the
mass, will be in general position dependent. Therefore, we believe
the information about the nontrivial boundary conditions or position
dependent backgrounds are carried by the full set of n-point
functions, the resulting counterterms, and the renormalized
parameters of the theory. Our preliminary investigations have
revealed that the main difference between our position dependent
counterterms and the free ones are maximal for positions which are
about a Compton wavelength away from the walls, although it is also
nontrivial at other places. Moreover, in the limit of large plate
separation our counterterms become position independent everywhere
except within one Compton wavelength away from the walls. In this
limit, the difference between our counterterms and the free ones,
before the transverse momentum integration, approaches a constant
value proportional to $1/2m$ away from the walls and $1/m$ at the walls. However
since these terms will multiply at least one-loop expressions which
are divergent, their difference will have significant consequences.
Here we use this procedure to compute the first-order radiation
correction to the Casimir energy for a real scalar field in 3+1
dimensions with $\phi^4$ self interaction. We compute this
correction for both a massive and a massless scalar fields and show
that the massless limit of the massive case exactly corresponds to
the massless case.

In addition, up to now most of the papers on the Casimir effect,
that we are aware of, use some from of \emph{analytic continuation}.
We share the point of view with some authors such as the ones in
\cite{kay,Mostepanenko} that the analytic continuation techniques
are not always completely justified physically. Moreover, like the
authors of the first of the aforementioned references, we have found
counterexamples, which we point out in this paper and elsewhere
\cite{moazzemi1,moazzemi2}. The counterexamples show that analytic
continuation techniques alone might not yield correct physical
results, and sometimes even give infinite results \cite{Bender}.
Therefore, we prefer to use a completely physical approach by
enclosing the whole system in a box of volume $V=L^3$, which
eventually can go to infinity, and calculating the difference
between the zero point energies of two different configurations. The
main idea of this method is actually due to T.H. Boyer
\cite{Boyer}, who used spheres instead of boxes. This we shall call
the ``box renormalization scheme" and can be used as a supplementary
part of other usual regularization or renormalization programs. This
box renormalization scheme, has the following advantages:
\begin{enumerate}
    \item Use of this procedure removes all of the ambiguities associated
    with the appearance of the infinities, and we use the usual prescription for
    removing the infinities in the regulated theory, as explained in Sec. \ref{massive case}. This is all done without
    resorting to any analytic continuation schemes.
    \item In order to calculate the Casimir energy we subtract
    two physical configuration of similar nature, e.g. both
    confined within finite regions, and not one confined and the other in
    an unbounded region.
    \item This method can be used as a check for the cases where
    analytic continuation yields finite results, and more
    importantly, can be used to obtain finite results when the
    former yields infinite results.
\end{enumerate}

More importantly, we have discovered for the case of parallel
plates, much to our surprise, that when the problem is set up
correctly, that is when proper counterterms are used, using analytic
continuation technique for NLO correction it seems impossible to
obtain finite results in any integer space-time dimensions, and
correct finite results in any non-integer space-time dimensions. We
will make this statement more explicit later on in Section
\ref{Analytic Continuation}.  However, this technique also gives the
correct results for the leading term for this geometry (see for
example \cite{Wolfram}). We should mention that some authors believe
that use of box regularization or renormalization procedures, in
which the size of the box eventually goes to infinities could be
avoided by using appropriate boundary conditions on the fields at
spatial infinity \cite{Nest}.

In Section 2 we calculate the leading order term for the Casimir
energy in $d$ space dimensional case. We do this first of all to
explain more completely the physical content of the problem and set
up our notations. Secondly this computation is just about as easy to
do in $d$ dimensions as  is in the three dimensional case. In
Section 3 we compute the first order radiative correction to this
energy. In order to do this we first state the renormalization
condition, and then derive an expression for the first order
radiative correction for a massive and massless scalar field. In
Section 4 we discuss the validity of the analytic continuation
techniques relevant to our problem. In Section five we give a brief
summary of our results and state our conclusions.

\section{The Leading Term of the Casimir Effect}\label{sce2}
The lagrangian density for a real scalar field with $\phi^4$
self-interaction is:
\begin{equation}\label{e1}
  {\cal L}(x)
  =\frac{1}{2}[\partial_{\mu}\varphi(x)]^{2}-\frac{1}{2}m_0^{2}\varphi(x)^{2}
  -\frac{\lambda_{0}}{4!}\varphi(x)^{4},
\end{equation}
where $m_{0}$ and $\lambda_{0}$ are the bare mass and bare coupling
constant, respectively. Here we calculate the leading term for the
Casimir energy in $d$ spatial dimensions. Obviously the leading
term, in contrast to the higher order corrections, is independent of
the form of the self-interaction. The Casimir energy is in general
equivalent to the work done on the system for bringing two parallel
plates from $\pm \infty$ to $\pm a/2$. As mentioned before, part of
our renormalization procedure is to enclose the whole system in a
$d$ dimensional cubical box of sides $L$. To compute this leading
term, we first compare the energies in two different configurations:
when the plates are at $\pm a/2$ as compare to $\pm b/2$. We name
the axis perpendicular to the plates the $z$ axis. To keep the
expressions symmetrical, we choose the coordinates so that the edges
of the confining box are at $\pm L/2$ in any direction.

\begin{figure}[th]
\begin{center} \includegraphics[width=7cm]{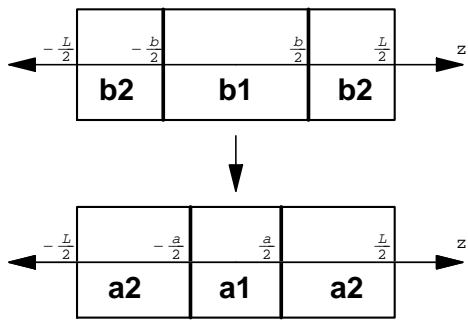}\caption{\label{fig:1} {\small
The geometry of the two different configurations whose energies are
to be compared. The labels a1, {\em etc.} denote the appropriate
sections in each configuration separated by the plates.}}
\label{geometry}
\end{center}
\end{figure}

The total zero point energy of the upper configuration in figure
(\ref{geometry}) will be called $E_{b}$ and of the lower one
$E_{a}$. In our box renormalization scheme we need to define the
Casimir energy as follows

\begin{equation}\label{e2:E.Cas.}
   E_{\mbox{\tiny \mbox{\tiny Cas.}}}=\lim_{b/a\rightarrow\infty}\left[\lim_{L/b\rightarrow\infty}
   \left(E_{a}-E_{b}\right)\right],
\end{equation}
where,
\begin{equation}\label{e3}
   E_{a}=E_{a_1}+2E_{a_2},\quad
   E_{b}=E_{b_1}+2E_{b_2}.
\end{equation}
Here we choose the Dirichlet boundary condition on all of the
boundaries. Then we can expand the field operator $\varphi$ in the
eigenstate basis appropriate to this boundary condition, and its
explicit second quantized form, for example in region $a_1$ becomes
\begin{eqnarray}\label{e4}
   \varphi_{_{a_1}}(x)&=&\int\frac{d^{d-1}\mathbf{k}^{\bot}}{(2\pi)^{d-1}}
   \sum_{n=1}^{\infty}\left(\frac{1}{a\omega_{a_1,n}}\right)^{1/2}
   \nonumber\\
   &&\hspace{-2cm} \times\Bigg\{e^{-i(\omega_{a_1,n}t-\mathbf{k^{\bot}}\textbf{.}
   \mathbf{x}^{\bot})}\sin\left[k_{a_1,n}(z+\frac{a}{2})\right]\textbf{a}_{n}\dd +e^{i(\omega_{a_1,n}t-\mathbf{k^{\bot}}\textbf{.}
   \mathbf{x}^{\bot})}\sin\left[k_{a_1,n}(z+\frac{a}{2})\right]
   \textbf{a}_{n}^{\dag}\Bigg\},
\end{eqnarray}
where,
\begin{eqnarray}\label{e5}
   \omega_{a_1,n}^{2}&=&m_0^{2}+{k^{\bot}}^{2}+k_{a_1,n}^{2},\quad
   k_{a_1,n}=\frac{n
   \pi}{a} \quad \mbox{and}\quad n=1,2,\ldots.
\end{eqnarray}
Here $\mathbf{k}^{\bot}$ and $k_{a_1,n}$ denote the momenta parallel
and perpendicular to plates, respectively. Also
$\textbf{a}_{n}^{\dag}$ and $\textbf{a}_{n}$ are creation and annihilation
operators  obeying the usual commutation relations:
\[\hspace{2.6cm}[\textbf{a}_{n},\textbf{a}^{\dag}
_{n^{_\prime}}]=\delta_{n,n^{_\prime}},\quad
[\textbf{a}_{n},\textbf{a}_{n^{_\prime}}]=[\textbf{a}^{\dag}_{n},
\textbf{a}^{\dag}_{n^{_\prime}}]=0,\]
 and $\textbf{a}|0\rangle=0$
defines the vacuum state in the presence of boundary conditions.
Using the above equations one can easily obtain
\begin{eqnarray}\label{e6}
   E^{(0)}_{a_1}=\dd\int d^{d}\textbf{x}\dd\langle0|{\mathcal H}^{(0)}|0\rangle
    &= &L^{d-1}\int\frac{d^{d-1}
   \mathbf{k}^{\bot}}{(2\pi)^{d-1}}\sum_{n=1}^{\infty}\frac{\omega_{a_1,n}}{2}\nonumber\\
   & =&\frac{L^{d-1}}{2}\frac{\Omega_{d-1}}{(2\pi)^{d-1}}
   \int^{\infty}_{0}dk k^{d-2}\sum_{n=1}^{\infty}\omega_{a_1,n},
\end{eqnarray}
where ${\mathcal H}^{(0)}$ denotes the usual free Hamiltonian
density, easily obtained from the Lagrangian density, and the
superscript $(0)$ denotes the zero (or leading) order term of this
energy. Also $k=|\mathbf{k}^{\bot}|$, and
$\Omega_d=\dd\frac{2\pi^{d/2}}{\Gamma(\frac{d}{2})}$ is the solid
angle in $d$-dimensions. Therefore,
\begin{equation}\label{e7}
   E^{(0)}_{a}-E^{(0)}_{b}=\frac{L^{d-1}}{2}\frac{\Omega_{d-1}}{(2\pi)^{d-1}}
   \int^{\infty}_{0}dk k^{d-2}\sum_{n}g(n),
\end{equation}
where,
\[g(n)=\omega_{a_1,n}+2\omega_{a_2,n}-\omega_{b_1,n}-2\omega_{b_2,n}.\]
Now we are allowed to use the Abel-Plana summation formula, since
 we now expect the summand to satisfy the strict conditions \cite{Henrici} for the validity of this formula.
 That is, we expect any reasonable renormalization program for calculating any measurable physical quantity to yield
finite results. The Abel-Plana summation formula gives
\begin{eqnarray}\label{e8:main}
   &&E^{(0)}_{a}-E^{(0)}_{b}=\frac{L^{d-1}}{2}\frac{\Omega_{d-1}}{(2\pi)^{d-1}}
   \int^{\infty}_{0}dk k^{d-2}
  \nonumber\\ &&\hspace{3cm}\times\left[\frac{-g(0)}{2}+\int^{\infty}_{0}g(x)dx+i
   \int^{\infty}_{0}\frac{g(it)-g(-it)}{\emph{e}^{2\pi
   t}-1}dt\right],
\end{eqnarray}
where $g(0)$  vanishes in this case due to our box renormalization.
The second term in the bracket, using suitable changes of variables,
becomes
\begin{eqnarray}\label{e9:sec.term}
   && \dd\frac{a}
   {\pi}\int^{\infty}_0d\kappa\left(m_0^2+k^2+\kappa^2\right)^{1/2}+2\frac{L-a}{2\pi}
   \int^{\infty}_0d\kappa\left(m_0^2+k^2+\kappa^2\right)^{1/2}\nonumber\\
 &&\quad\dd-\frac{b}{\pi}\int^{\infty}_0d\kappa
   \left(m_0^2+k^2+\kappa^2\right)^{1/2}-2\frac{L-b}{2\pi}\int^{\infty}_0d\kappa
   \left(m_0^2+k^2+\kappa^2\right)^{1/2}=0,
\end{eqnarray}
where $\kappa$ for example in the first term denotes $\dd\frac{
n\pi}{a}$ treated as a continuous variable. The above calculation
shows that this term is exactly zero. Therefore, only the branch-cut
term (the last term in eq.~(\ref{e8:main})) gives nonzero
contribution and the final result is
\begin{eqnarray}\label{e10}
     \hspace{-.8cm}E^{(0)}_{a}-E^{(0)}_{b}&=&-\frac{2L^{d-1}m_0^{^{(d+1)/2}}}{(4\pi)^{(d+1)/2}}
    \dd \sum^{\infty}_{j=1}\frac{1}{j^{(d+1)/2}}
    \Bigg\{\frac{K_{_{(d+1)/2}}(2ajm_0)}{a^{(d-1)/2}}-\frac{K_{_{(d+1)/2}}
   (2bjm_0)}{b^{(d-1)/2}}\nonumber\\&&+\frac{2K_{_{(d+1)/2}}[(L-a)jm_0]}{(\frac{L-a}{2})^{^{(d-1)/2}}}
   -\frac{2K_{_{(d+1)/2}}[(L-b)jm_0]}{(\frac{L-b}{2})^{^{(d-1)/2}}}\Bigg\},
\end{eqnarray}
where $K_n(x)$ denotes the modified Bessel function of order $n$.
Using eq.~(\ref{e2:E.Cas.}) for the Casimir energy and noting that
$K_n(x)$ is strongly  damped as $x$ goes to infinity, only the first
term remains when the limits are taken, and the result is
\begin{equation}\label{e11:res.z.term}
   E^{(0)}_{_{\mbox{\tiny Cas.}}}=-\frac{2L^{d-1}}{(4\pi)^{(d+1)/2}}
   \frac{m_0^{(d+1)/2}}{a^{(d-1)/2}} \sum^{\infty}_{j=1}
   \frac{K_{_{(d+1)/2}}(2ajm_0)}{j^{(d+1)/2}}.
\end{equation}
This is the result for the leading term for the Casimir energy in
$d$-dimensions, on which all authors agree (see for example
\cite{Wolfram,Svaiter}). It is important to note that we, unlike
most other authors, derived this result without any use of analytic
continuation. If we set $d=3$, we have
\begin{equation}\label{e12}
   E^{(0)}_{_{\mbox{\tiny Cas.}}}=-\frac{L^2m_0^2}{8\pi^2a}\sum^{\infty}_{j=1}
   \frac{K_2(2ajm_0)}{j^2},
\end{equation}
with the following limits,
\begin{equation}\label{e13}
   E^{(0)}_{_{\mbox{\tiny Cas.}}}\rightarrow\left\{
   \begin{array}{ll}
        \dd\frac{-L^2}{8\pi^2a}\sum_{j}\frac{1}{2a^2j^4}
   =\dd\frac{-L^2\pi^2}{1440a^3} & \quad\mbox{as}\quad m_0\rightarrow0, \\
         \raisebox{-5mm}{$\dd\frac{-L^2}{8\sqrt{2}}(\frac{m_0}{\pi a})^{3/2}e^{-2am_0}$} &
         \raisebox{-5mm}{$\quad\mbox{as} \quad am_0\gg1.$}
   \end{array}\right.
\end{equation}
The results are in agreement with what exists in literature (see for
instance \cite{Svaiter,Mostepanenko}). It is interesting to note
that for the massless case, the result is, not surprisingly, exactly
half of the corresponding expression for the electromagnetic case.

\section{First-Order Radiative Correction}

Now we calculate the next to the leading order (two loop quantum
correction) shift of the Casimir energy for a scalar field in
$\phi^{4}$ theory using the renormalized perturbation theory in
$3+1$ dimensions. As mentioned before, the main idea of our work is
that when a systematic treatment of the renormalization program is
done, the counterterms needed to retain the renormalization
conditions, automatically turn out to be position dependent. This,
as we shall see, will have profound consequences. However, our main
scheme of canceling the divergences  using counterterms and a few
input experimental parameters, is in complete  conformity with the
standard renormalization approach. To set the stage for the
calculations, we shall very briefly state the renormalization
procedure and conditions.
\subsection{Renormalization Conditions}
The $\phi^{4}$ Lagrangian eq.(\ref{e1}), after rescaling the field
$\varphi=Z^{1/2}\varphi_{r}$, where $Z$ is called the field strength
renormalization, and the standard procedure for setting up the
renormalized perturbation theory, becomes (see for example
\cite{Peskin}),
\begin{eqnarray}\label{e17}
   &&{\cal L}(x)=\frac{1}{2}[\partial_{\mu}\varphi_{r}(x)]^{2}-\frac{1}{2}m^{2}
   \varphi_{r}(x)^{2}-\frac{\lambda}{4!}\varphi_{r}(x)^{4}\qquad\nonumber\\&&\hspace{2cm}+
   \frac{1}{2}\delta_{Z}[\partial_{\mu}\varphi_{r}(x)]^{2}
   -\dd\frac{1}{2}\delta_{m}\varphi_{r}(x)^{2}-\dd\frac{\delta_{\lambda}}{4!}\varphi_{r}(x)^{4},
\end{eqnarray}
where $\delta_{m},\delta_{\lambda},\delta_{Z}$ are the counterterms,
and $m$ and $\lambda$ are the physical mass and physical coupling
constant, respectively. In this problem we are to impose boundary
conditions on the field at the walls. An alternative approach would
be to add appropriate external potentials to the Lagrangian so as to
maintain the boundary conditions on the fields \cite{Graham2}. We
will use the first approach. Obviously the presence of nontrivial
boundary conditions breaks the translational invariance and hence
momenta will no longer be good quantum numbers. Therefore we find it
easier to impose the renormalization conditions in the configuration
space. For example, the standard expression for the two-point
function
is,
\begin{eqnarray}\label{2pfun.}
    \langle\Omega
    |T\{\phi(x_1)\phi(x_2)\}|\Omega\rangle&=&\lim_{T\to\infty(1-i\epsilon)}
    \frac{\langle0|\int{\cal D}\phi\phi(x_1)\phi(x_2)e^{i\int_{-T}^T
    {\cal L} d^4x }\}|0\rangle}{\langle0|\int{\cal D}\phi e^{i\int_{-T}^T {\cal L}
    d^4x}|0\rangle}.
\end{eqnarray}
Since the birth of quantum field theory, as far as we know, the
assertion has always been that the above expressions can be expanded
systematically when the problem is amenable to perturbation theory.
For example, in the context of renomalized perturbation theory, as
indicated in eq.(\ref{e17}), we can symbolically represent the first
few terms of the perturbation expansion of eq.(\ref{2pfun.}) by
\begin{equation}\label{e18:renor.con1.}
   \raisebox{-4.5mm}{\includegraphics{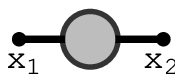}}=\raisebox{-3.3mm}{\includegraphics{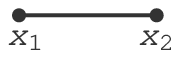}}
   +\raisebox{-3.7mm}{\includegraphics{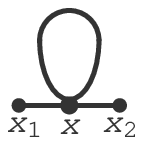}}+\raisebox{-3.9mm}{\includegraphics{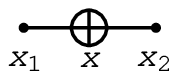}}+\dots.
\end{equation}
where \raisebox{-3mm}{\includegraphics{16}} refers to the
appropriate counterterm. It is obvious that the above expression
represents a systematic perturbation expansion, and most
importantly, all of the propagators on the right hand side should be
the one appropriate to the problem under consideration, that is they
should have the same overall functional form as the first term. Our
first renormalization condition is that the renormalized mass $m$
should be the pole of the propagator represented by the first term
in (\ref{e18:renor.con1.}). This implies the second and third
diagrams should cancel each other out in the lowest order, and this
in turn implies the cancelation of the UV divergences in that order,
and that the counterterms will in general turn out to be position
dependent. The renormalized mass $m$ will then naturally turn out to
be position dependent as well. However, we only need to fix the
value of $m(x)$ at one position between the plates by our
renormalization condition. The exact functional dependence of $m(x)$
will then be completely determined by the theory. That is, we insist
the overall structure of the renormalization conditions such as
above, and the counterterms appearing in them should be determined
solely from within the theory, and not for example be imported from
the free case. The equations are self deterministic and there is no
need to take such actions. Obviously we still need a few
experimental input parameters for the complete renormalization
program, such as $m(x)$ for some $x$. Analogous expression and
reasonings could be easily stated for the four-point function.

To one-loop order, the standard renormalization conditions applied
to eq.~(\ref{e18:renor.con1.}) and its four-point counterpart, give
\begin{equation}\label{e20:counterterms}
   \delta_{Z}(x)=0,\quad\delta_{m}(x)=\frac{-i}{2}\raisebox{-2.4mm}{\includegraphics{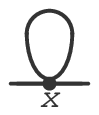}}
   =\frac{-\lambda}{2}G(x,x); \quad\mbox{and}
   \quad\delta_{\lambda}(x)=0,
\end{equation}
respectively. Here $G(x,x')$ is the propagator of the real scalar
field and $x=(t,\mathbf{x})$. Obviously the counterterms
automatically incorporate the boundary conditions and are position
dependent, due to the dependence of the two and four-point functions
on such quantities. Now, the higher order contributions to the vacuum energy in the
interval a1 (i.e. $z \in [\frac{-a}{2},\frac{a}{2}]$) are
\begin{eqnarray}\label{e21:vacc-pol}
   \Delta E_{a_1}= E^{(1)}_{a_1}+ E^{(2)}_{a_1}+\dots= \int_{V}
   d^3{\bf x}\langle\Omega|{\cal H}_{_I}|\Omega\rangle\qquad\nonumber\\=i\int_{V} d^3{\bf x}\left(\frac{1}{8}
   \raisebox{-7mm}{\includegraphics{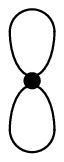}}\ +\frac{1}{2}
   \raisebox{-1mm}{\includegraphics{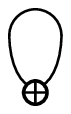}} \ +\frac{1}{8}\raisebox{-7mm}{\includegraphics{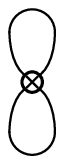}}+\dots
   \right),
\end{eqnarray}
where\raisebox{-3mm}{\includegraphics{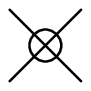}}$=-i\delta_{\lambda}(x)$
and
\raisebox{-1mm}{\includegraphics{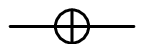}}$=i[p^{2}\delta_{Z}(x)-\delta_{m}(x)]$
refer to the counterterms. Accordingly, the ${\cal O}(\lambda)$
contribution to the vacuum energy is
\begin{eqnarray}\label{e22:vacc-ene}
   E^{(1)}_{a_1}&=& i\int_{V} d^3{\bf x}\left(\frac{1}{8}
   \raisebox{-7mm}{\includegraphics{8}}\ +\frac{1}{2}
   \raisebox{-1mm}{\includegraphics{9}} \ \right)\hspace{3cm}\nonumber\\&=&i\int_{V}
   d^3{\bf x}\left[\frac{-i\lambda}{8}G^{2}_{a_1}(x,x)
   -\frac{i}{2}
    \delta_{m}(x)G_{a_1}(x,x)\right],
\end{eqnarray}
where $G_{a_1}(x,x')$ is the propagator of the real scalar field in
region a1. Using eqs.(\ref{e20:counterterms}) and
(\ref{e22:vacc-ene}), we obtain,
\begin{eqnarray}\label{e23:E.Corr.}
   E^{(1)}_{a_1}=\dd\frac{-\lambda}{8}\int_{V}G^{2}_{a_1}(x,x)d^3{\bf x}.
\end{eqnarray}
\subsection{The Massive Case}\label{massive case}
As mentioned before, here we choose the Dirichlet boundary condition
on the plates. Then, after the usual wick rotation, the expression
for the Green's function in the four dimensional Euclidean space
becomes
\begin{eqnarray}\label{EucGreesfun.}
  &&\hspace{-.7cm}G_{a_1}(x,x')=\frac{2}a\int\frac{d^{3} k}{(2\pi)^{3}} \sum_n
\frac{e^{-\omega(t-t')}e^{-i{\small\bf
k^{\bot}}.({\bf{x}}^{\bot}-{\bf{x'}}^{\bot})}
  \sin\left[k_{a1,n}(z+\frac{a}{2})\right]
\sin\left[k_{a1,n}(z'+\frac{a}{2})\right]}{k^2+k_{a1,n}^2+m^2+i\epsilon}.\nonumber\\
\end{eqnarray}
It is very important to note that the $\dd\delta_{m}(x)
=\frac{-\lambda}{2}G(x,x)$ (c.f. eq. (\ref{e20:counterterms})) is
explicitly position dependent. Using eq.~(\ref{EucGreesfun.}) and
eq.~(\ref{e23:E.Corr.}) and carrying out the spatial integration one
obtains,
 \begin{eqnarray}\label{Einf.}
  E_{a_1}^{(1)}=-\frac{\lambda L^2}{32\pi^4a}\Bigg[\sum_{n,n'}\int_0^\infty dkk^2
\frac{1}{k^2+k_{a1,n}^2+m^{2}} \int_0^\infty dkk^2
\frac{1}{k^2+k_{a1,n'}^2+m^{2}}\nonumber\\+\frac{1}2
\sum_{n}\left(\int_0^\infty dkk^2
\frac{1}{k^2+k_{a1,n}^2+m^{2}}\right)^2\Bigg].
\end{eqnarray}
To compute the integrals we regulate them by a momentum cutoff
$\Lambda$. Then we have
\begin{eqnarray}\label{E1}
 E_{a_1}^{(1)} &=&-\frac{\lambda
L^2}{32\pi^4a}\lim_{\Lambda_{a_1,n}\rightarrow\infty}\Bigg[\sum_{n,n'}\left(\Lambda_{a_1,n}^2 - \pi\omega'
_{a1,n}\Lambda_{a_1,n}+\frac{\pi^2}4\omega' _{a1,n}\omega'
_{a1,n'}\right)\nonumber
\\&&\hspace{2cm}+\frac{1}2\sum_n\left(\Lambda_{a_1,n}^2-\pi\omega'
_{a1,n}\Lambda_{a_1,n}+\frac{\pi^2}4 \omega'
_{a1,n}{^2}\right)\Bigg],
\end{eqnarray}
where $\omega' _{a1,n}{^2}=k_{a1,n}^2+m^2$. Now the NLO for the
Casimir energy, eq.(\ref{e2:E.Cas.}), has four terms the first one
of which is shown above. It is clear that we can adjust the cutoffs
for each region so that the quadratic divergences coming from
different regions cancel each other. Similar cancelation occurs for
the linear divergences. The cancelation of these divergent
quantities without any residual finite terms is the usual practice
in regulated theories \cite{Mostepanenko}, and this is the
prescription that we shall use. We shall comment on this further
when we use it one more time for eq.(\ref{e27:corr.cas.E}) below.
Hence we get
\begin{eqnarray}\label{Ea1wEb1w}
&&E^{(1)}_{a}-E_{b}^{(1)} =-\frac{\lambda
L^2}{128\pi^2}\sum_{n}\Bigg[\sum_{n'}\left(\frac{\omega'
_{a1,n}\omega' _{a1,n'}}a -\frac{\omega' _{b1,n}\omega'
_{b1,n'}}b+4\frac{\omega' _{a2,n}\omega' _{a2,n'}}{L-a}
-4\frac{\omega' _{b2,n}\omega'
_{b2,n'}}{L-b}\right)\nonumber\\&&\hspace{5cm}+\frac{1}2\left(
\frac{\omega _{a1,n}'^2}a-\frac{\omega _{b1,n}'^2}b +4\frac{\omega
_{a2,n}'^2}{L-a}-4\frac{\omega _{b2,n}'^2}{L-b}\right)\Bigg].
\end{eqnarray}
This computation is obviously complicated and plagued with a
multitude of infinities. As explained before using the usual
renormalization programs in conjunction with our box renormalization
scheme, should eliminate all of the infinities, as might be apparent
from the above equation. However, proper regularization schemes
should still be implemented and proper care taken when handling
these infinite expressions. All the summations appearing in the
eq.~(\ref{Ea1wEb1w}) are separately infinite. We want to use the
Abel-Plana formula to convert these sums into integrals. However
these sums do not satisfy the stringent requirements stated in the
Abel-Plana theorem  for such a conversion \cite{Henrici,Saharian}.
However our box renormalization scheme provides a solution by
subtracting these double sums as indicated in eq.~(\ref{e2:E.Cas.}).
Now we can expect this new summand to satisfy the requirements for
the Abel-Plana theorem. Then all the infinities actually cancel and
the result for the two-loop correction reduces to (see Appendix for
more details):
\begin{eqnarray}\label{e27:corr.cas.E}
     &&\hspace{-1.2cm}E^{(1)}_{a}-E^{(1)}_{b}=\frac{-\lambda L^2}{128\pi^2}
    \Bigg[f(a)-f(b)+2f(\frac{L-a}{2})-2f(\frac{L-b}{2})\nonumber\\&&\hspace{1.2cm}
    +\underbrace{\frac{2m^2}{\pi}\left(B(a)-B(b)
   +2B(\frac{L-a}{2})-2B(\frac{L-b}{2})\right)
   \int_{0}^{\infty}ds\sqrt{1+s^{2}}}_{\dd={\cal R}}\Bigg],
\end{eqnarray}
where $\dd f(a)=B(a)\left(\frac{B(a)}{a}-\frac{m}{a}\right) $ and
$B(a)$, defined by the following expression
\begin{eqnarray}\label{eA4:diff.B}
   &&\hspace{-.8cm} B(a)=\dd
-\frac{2am^2}{\pi}\int_{1}^{\infty}\frac{\sqrt{t^2-1}}{e^{2amt}-1}dt
=-\frac{m}{\pi }\sum_{j=1}^{\infty}\frac{K_{1}(2amj)}{j},
\end{eqnarray}
refers to the so called branch-cut term of the Abel-Plana summation
formula and is a finite quantity. Note that the last integral in
eq.~(\ref{e27:corr.cas.E}) seems to diverge so it must be properly
regularized. We prefer to use a regularization scheme for this
integral term which is analogous to the zeta function regularization
for the sums. That is, we set the power of the integrand to
$\frac{d}{2}-1$ and in the final stage we let $d$ approach three.
Henceforth we shall also refer to this as zeta function
regularization. We should note that this regularization is not
sufficient to avoid the use of analytic continuation. However, as we shall see, by
adding an auxiliary cutoff regulator we can
ultimately avoid resorting to any analytic continuation. We thus
obtain
 \begin{equation}  \label{a}
\hspace{-.5cm}\int_{0}^{\infty}(1+\kappa^{2})^{\frac{d}{2}-1}d\kappa\quad\to\quad\dd
\int_{0}^{K}(1+\kappa^{2})^{\frac{d}2-1}d\kappa =K \
{_2F_1}\left(\frac{1}2,1-\frac{d}2,\frac{3}2,-K^2\right).
\end{equation}
The asymptotic behavior of the hypergeometric function
${_2F_1}\left(\frac{1}2,1-\frac{d}2,\frac{3}2,-K^2\right)$
multiplied by $K$ for the large $K$ is,
\begin{eqnarray}  \label{A6}
\hspace{-.8cm}K \
{_2F_1}\left(\frac{1}2,1-\frac{d}2,\frac{3}2,-K^2\right) &{\buildrel
{K \to \infty } \over \longrightarrow }& \quad \frac{\sqrt
\pi\Gamma(\frac{1-d}2)}{2\Gamma(1-\frac{d}2)}+\frac{1}{d-1}K^{d-1}
+\frac{d-2}{2(d-3)}K^{d-3}+\dots \nonumber\\
&{\buildrel {d \to 3 } \over \longrightarrow }& \quad
\frac{1}{2}\left[K^{2}+\ln(K)+\ln(2)+1/2\right].
\end{eqnarray}
Referring to eq. (\ref{e27:corr.cas.E}) and adhering to the usual
prescription described for eq. (\ref{E1}), we can cancel the
quadratic and logarithmic divergences using our full freedom to
choose four different cutoffs in the four distinct integration
regions corresponding to \{a1, a2, b1, b2\} \footnote{One may argue
that ambiguities always exist in problems where one has to subtract
infinite quantities, and the Casimir problems certainly fall into
this category. Two methods are in common use: First is the analytic
continuation techniques which, although usually yield correct
results, do not have a very solid physical justification and also
sometimes yield infinite results. We have devoted Section 4 to this
subject. Second is the regularization schemes, which is what we have
used. In the latter category when the problem is regularized, one
can make a systematic expansion of the quantities in question in
terms of the regulators. Then the terms which tend to infinity when
the regulators are removed and the finite terms naturally appear
separately. See for example eqs. (\ref{E1},\ref{A6}). What is almost
invariably done is to adjust the regulators so that the singular
terms exactly cancel each other, i.e. without  extracting any extra
finite piece from the difference between the infinite quantities
(see for example \cite{Mostepanenko}). This is also apparent in the
leading term for the Casimir energy in eq. (\ref{e9:sec.term})
where, as explained in the Appendix, The four changes of variables
are equivalent to choosing four different cutoffs. One could have
adjusted them so that as usual the infinities cancel, but any finite
term would remain. However, the well known answer is obtained only
when there is no remaining extra finite term in this subtraction
scheme. This is the prescription that we have used. However, we do
believe that this is a subject that needs further study.}. Hence we
get,
\begin{equation}  \label{calR}
 \dd {\cal R}=\frac{m^2(\ln2+1/2)}{\pi}
\left(B(a)-B(b)+2B(\frac{L-a}{2})-2B(\frac{L-b}{2})\right).
\end{equation}
Using eqs.~(\ref{e2:E.Cas.}), (\ref{e27:corr.cas.E}),
(\ref{eA4:diff.B}) and (\ref{calR}), we obtain
\begin{equation}\label{fin.res.}
   \hspace{-.7cm}E^{(1)}_{_{\mbox{\tiny Cas.}}}=\frac{-\lambda L^2m}{128\pi^3}
   \sum_{j=1}^{\infty}\frac{K_{1}(2amj)}{j}\left[\frac{m}{\pi a}
    \sum_{j'=1}^{\infty}\frac{K_{1}(2amj')}{j'}+\frac{m}{a}-\frac{m^2}{\pi}(\ln2+1/2)
   \right].
\end{equation}
This is the two-loop radiative correction to the Casimir energy.
This result is obviously finite, and we believe it could not have
been obtained with any regularization or analytic continuation
schemes in common use. Our result defers from \cite{Baron1, Caval}.

\begin{figure}[th]\begin{center} \includegraphics[width=11cm]{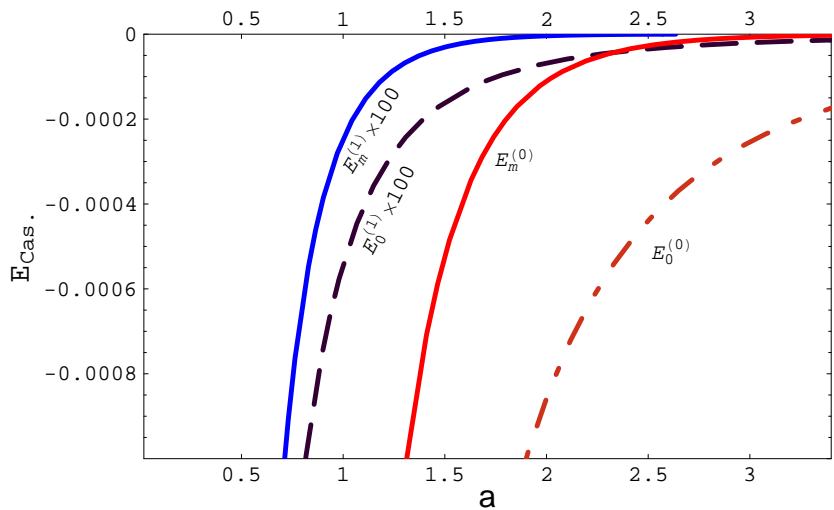}
\caption{\label{mfd3}\small The leading terms for the Casimir energy
and its NLO corrections as a function of $a$ for the massive ($m=1$)
and massless cases for $\lambda=0.1$. The correction term for the
massless case is always negative. However the correction term for
the massive case has a peak of magnitude $\simeq +1.57\times
10^{-9}$ at $a\simeq 3.04 $ and asymptotes to zero from above.}
\end{center}\end{figure}

Two particular limits are interesting to calculate; The large mass
limit, $ma\gg1$, and small mass limit, $m\to 0$. In these limits
eq.(\ref{fin.res.}) becomes
\begin{equation}
 \left\{
   \begin{array}{ll}
     \dd  E^{(1)}_{_{\mbox{\tiny Cas.}}}\quad{\buildrel {am\gg1 } \over
 \longrightarrow }\quad
\frac{\lambda
L^2m^2}{256\pi^3}\frac{\ln2+1/2}{\sqrt{\pi}}\sqrt{\frac{m}{a}}
   e^{-2am},
     &  \\
     \raisebox{-9mm}{$\dd E^{(1)}_{_{\mbox{\tiny Cas.}}}
   \quad {\buildrel {m\to0 } \over
 \longrightarrow }\quad -\frac{\lambda L^2}{512\pi^4a^3}
   \left(\sum_{j=1}^{\infty}\frac{1}{j^2}\right)^2=-\frac{\lambda L^2}{18432a^3}.$}
     &
   \end{array}\right.
\end{equation}
Our massless limit differs from the analogous results that can be
extracted from refs. \cite{Baron1, Caval,Symanzik,Krech} by a minus
sign. Figure \ref{mfd3} illustrates the graphs for the leading and
the NLO terms for the Casimir energy as a function of $a$, in the
massive case and its massless limit.

\subsection{The Massless Case}
In the massless case it is sufficient to set the pole of the
propagator to zero, i.e. one can set $m=0$ in the
eq.~(\ref{EucGreesfun.}), hence in the eq.(\ref{Ea1wEb1w}) we have,
for example $\omega_{a_1,n}'=k_{a_1,n}$. Accordingly this equation
becomes
\begin{eqnarray}\label{Ea1w..}
\hspace{-.5cm}E^{(1)}_{a}-E_{b}^{(1)}& =& \frac{-\lambda
L^2}{128}\sum_{n}\Bigg[\sum_{n'} \left(\frac{1}{a^3}nn'
-\frac{1}{b^3}nn' +\frac{16}{(L-a)^3}nn'
-\frac{16}{(L-b)^3}nn'\right) \nonumber\\\noalign{\vspace{.2cm}}
&&\hspace{2cm}+\frac{1}{2a^3}n^2-\frac{1}{2b^3}n^2+\frac{8}{(L-a)^3}n^2
-\frac{8}{(L-b)^3}n^2\Bigg].
\end{eqnarray}
We use Abel-Plana summation formula to transform the sums into
integrals, then one can remove the divergent integrals by changing
variables as explained before. Then the result is
\begin{eqnarray}\label{Ea...}
 E^{(1)}_{a}-E_{b}^{(1)}
&=&-\frac{\lambda L^2}{18432} \left(\frac{1}{a^3}
-\frac{1}{b^3}+\frac{16}{(L-a)^3}-\frac{16}{(L-b)^3}\right).
\end{eqnarray}
As before the eq.(\ref{e2:E.Cas.}) gives the Casimir energy,
\begin{equation}\label{E.Cas.d=3}
  \dd E^{(1)}_{\mbox{\tiny \mbox{\tiny Cas.}}}
=-\frac{\lambda L^2}{18432a^3}.
\end{equation}
This result is in exact agreement with the small mass limit
calculated in previous subsection.

Note that we are explicitly assuming that $\delta_{m=0}(x) \neq 0$.
we like to stress that this quantity should not in general be a
priori set to zero. This is in contrast to the view expressed in for
example refs.\cite{Symanzik,Krech,Ritberg, Geresd}. This is yet
another important counterexample for the validity of analytic
continuation: As is well known the massless limit of the analytic
continuation of the mass counterterm in $\phi^4$ theory is zero for
space-time dimensions bigger than two. However one cannot
renormalize the massless theory without the mass counterterm (see
for example \cite{Peskin}).

\section{On the Analytic Continuation Techniques}\label{Analytic Continuation}
In the previous section we saw that enclosing the whole system in a
box is very useful for removing all of the infinities which appear
in the calculations. In this section, for comparison purposes, we
repeat some of those steps using analytic continuation techniques.
We can then show or point out three examples and two counterexamples
for the validity of the analytic continuation techniques.

The first example for the validity of these techniques is the
derivation of the leading term mentioned in the Section \ref{sce2}. The
second example is the following: Starting with
eq.~(\ref{EucGreesfun.}) which is obviously infinite, one can
regularize it by dimensional regularization:
\begin{equation}\label{Gr.A.C.}
G_{a_1}(x,x)=\frac{2\Gamma(1-d/2)}{(4\pi)^{d/2}a}
\sum_{n}\omega_{a_1,n}'^{d-2}\sin^2\left[k_{a_1,n}(z+\frac{a}{2})\right].
\end{equation}
This equality holds only for $d<1$, however it can be analytically
continued to $d\ge1$. Putting this Green's function in
eq.~(\ref{e23:E.Corr.}) and performing the space integration and
repeating analogous steps for the other regions, one easily obtains:
\begin{eqnarray}\label{Ea1wEb1w2...}
&&E^{(1)}_{a}-E_{b}^{(1)} =\nonumber\\&&-\frac{\lambda
L^{d-1}\Gamma^2(1-d/2)}{8(4\pi)^{d}}\sum_{n}\Bigg[\sum_{n'}\left(\frac{\omega'^{d-2}
_{a1,n}\omega'^{d-2} _{a1,n'}}a -\frac{\omega'^{d-2}
_{b1,n}\omega'^{d-2}_{b1,n'}}b+4\frac{\omega'^{d-2}
_{a2,n}\omega'^{d-2} _{a2,n'}}{L-a} -4\frac{\omega'^{d-2}
_{b2,n}\omega'^{d-2}
_{b2,n'}}{L-b}\right)\nonumber\\&&\hspace{4.5cm}+\frac{1}2\left(
\frac{\omega _{a1,n}'^{2d-4}}a-\frac{\omega _{b1,n}'^{2d-4}}b
+4\frac{\omega _{a2,n}'^{2d-4}}{L-a}-4\frac{\omega
_{b2,n}'^{2d-4}}{L-b}\right)\Bigg].
\end{eqnarray}
Putting $d=3$ in eq.~(\ref{Ea1wEb1w2...}), eq.~(\ref{Ea1wEb1w}) is
obtained. Consequently we have shown that the analytic
continuation for the Green's function, in this case, gives the
correct result. The third example for the validity of the analytic
continuation technique is that with its mere use one can obtain
eq.~(\ref{Ea...}) from eq.~(\ref{Ea1w..}).

The first counterexample can be illustrated in the continuation of
derivation of the NLO Casimir energy from eq.~(\ref{Ea1wEb1w2...}).
Using the Abel-Plana formula we encounter the following term
\begin{eqnarray}
\int_{0}^{\infty}(1+\kappa^{2})^{\frac{d}2-1}d\kappa =\frac{\sqrt
\pi\Gamma(\frac{1-d}2)}{2\Gamma(1-\frac{d}2)},
\end{eqnarray}
and this leads to the a term which contains $\dd
\Gamma(1-d/2)\Gamma(\frac{1-d}2)$. This term is infinite in any
integer space dimension. This is an obvious counterexample for the
use of this analytic continuation technique. As for the second
counterexample, it concerns the massless limit of the analytic
continuation of the mass counterterm, as explained at the end of the
previous subsection.

\section{Conclusions}
We have introduced a new concept in this paper. We have insisted
that the renormalization program should completely and
self-consistently take into account the boundary conditions or any
possible nontrivial backgrounds which break the translational
invariance of the system. We have shown that the problem is
self-contained and the above program is accomplishable. To be more
specific, there should be no need to import counterterms from the
free theory, or even supplementing them with the \emph {ad hoc}
attachment of extra surface terms, to remedy the divergences
inherent in this theory. In general this breaking of the
translational invariance is reflected in the nontrivial position
dependence of all the n-point functions. As we have shown, this has
profound consequences. For example in the case of renormalized
perturbation theory, the counterterms and hence the radiative
corrections to parameters of the theory, i.e. $m$ and $\lambda$,
automatically turn out to be position dependent. In this regard we
disagree with the authors who use the free counterterms (see the
Introduction for actual references). Obviously we still need a few
experimental input parameters for the complete renormalization
program, such as $m(x)$ for some $x$. However, the interesting point
is that the theory then completely determines $m(x)$.

Secondly we have used a supplementary renormalization scheme to be
used along side the usual renormalization program. In computations
of these sorts, there usually appears infinities which can sometimes
be removed by the usual renormalization programs that often contain
some sort of analytic continuation. These procedures are sometimes
ambiguous. Our scheme is simply to confine the whole physical system
in a box, and to compute the difference between the values of the
physical quantity in question in two different configurations. Use
of this procedure removes all of the ambiguities associated with the
appearance of the infinities, and we use the usual prescription for
removing the infinities in the regulated theory. Using our method,
we have computed the zero and first order radiative correction to
the Casimir energy for the massive and massless real scalar field in
3+1 dimensions. For the zero order, our results are identical with
what exists in the literature. However, our first order results are
markedly different from those reported in refs.\cite{Baron1,Caval}.
Our results for the massive and massless cases are different from
theirs due to the aforementioned conceptual differences. As we have
shown our results for the massless case and the massless limit of
the massive case are identical. However, the massless limit of their
massive results agrees with the ``exact" results obtained in
\cite{Symanzik,Krech}, who set $\delta_{m}$ equal to zero in their
massless cases, or their equivalent. This is our second main
difference in approach to the problem. As mentioned before, we
believe that $\delta_{m}$ should not be arbitrarily set to zero even
in the massless case, since in that case the renormalization
conditions can no longer be fully implemented, although the theory
is still in principle renormalizable. In this regards, we like to
emphasize that their results for massless case is infinite in 1+1
dimensions.

\vskip20pt\noindent {\large {\bf
Acknowledgements}}\vskip5pt\noindent It is a great pleasure for us
to acknowledge the useful comments of Michael E. Peskin which
helped us resolve one of the main difficulties in our calculations.
We would also like to thank the Referee for illuminating comments.
This research was supported by the office of research of the Shahid
Beheshti University. R.M. would also like to thank the Qom
University for financial support. \vskip10pt

\appendix
\section{Use of the Abel-Plana summation formula in our subtraction scheme} \label{Calculation}

In this appendix we present the details of the calculations leading
to eq.~(\ref{e27:corr.cas.E}). The Abel-Plana summation formula (see
for example \cite{Saharian}) is:
\begin{equation}\label{eA1:abel.plana}
    \sum_{n=1}^{\infty}f(n)=-\frac{f(0)}{2}+\int_{0}^{\infty}f(x)dx
    +i\int_{0}^{\infty}\frac{f(it)-f(-it)}{e^{2\pi t}-1}dt.
\end{equation}

In order to obtain first-order radiative correction, just as we
discussed in Section \ref{sce2} for leading term, we need to compute
$E^{(1)}_{a}-E^{(1)}_{b}$. We rewrite the eq.~(\ref{Ea1wEb1w}) as,
\begin{eqnarray}\label{A2}
E^{(1)}_{a}&-&E_{b}^{(1)} =\frac{-\lambda
L^2}{128\pi^2}\sum_{n}\Bigg[\sum_{n'}
\Bigg(\frac{1}{a}S(a,n)S_{a_1}(n') -\frac{1}{b}S(b,n)S(b,n')
\nonumber\\
& +&\frac{4}{L-a}S(\frac{L-a}2,n)S(\frac{L-a}2,n')
-\frac{4}{L-b}S(\frac{L-b}2,n)S(\frac{L-b}2,n')\Bigg)\nonumber\\
& +&\frac{S^2(a,n)}{2a}-\frac{S^2(b,n)}{2b}
+\frac{2S^2(\frac{L-a}2,n)}{L-a}
-\frac{2S^2(\frac{L-b}2,n)}{L-b}\Bigg],
\end{eqnarray}
where $S(a,n)=\left(m^2+\frac{n^2\pi^2}{a^2}\right)^{1/2}$. Using
the Abel-Plana formula eq.~(\ref{eA1:abel.plana}), we obtain
\begin{eqnarray}\label{E1asE1bs}
\hspace{-.5cm}E^{(1)}_{a}&-&E_{b}^{(1)}=\nonumber\\&&\hspace{-1cm}\frac{-\lambda
L^2}{128\pi^2}\sum_{n}
\Bigg(\frac{m}{a}S(a,n)-\frac{m}{b}S(b,n)+\frac{4m}{L-a}S(\frac{L-a}2,n)
-\frac{4m}{L-b}S(\frac{L-b}2,n)\nonumber\\
& +&\frac{1}{\pi}\int_{0}^{\infty}\sqrt{m^{2}+s'^{2}}ds'
\left[S(a,n)-S(b,n)+2S(\frac{L-a}2,n)-2S(\frac{L-b}2,n)\right]\nonumber\\
& +&\frac{B(a)S(a,n)}{a} -\frac{B(b)S(a,n)}{b}
+\frac{4B(\frac{L-a}{2})S(a,n)}{L-a}
-\frac{4B(\frac{L-b}{2})S(a,n)}{L-b}\nonumber\\
& +&\frac{S^2(a,n)}{2a}-\frac{S^2(b,n)}{2b}
+\frac{2S^2(\frac{L-a}2,n)}{L-a}
-\frac{2S^2(\frac{L-b}2,n)}{L-b}\Bigg),
\end{eqnarray}
Using 
eq.~(\ref{eA1:abel.plana}) again, and making appropriate changes of
variables to make the integrals dimensionless, all the actual
infinities cancel and we finally obtain,

\begin{eqnarray}  \label{Res.d=3..}
  && E^{(1)}_{a}-E^{(1)}_{b}=\nonumber\\
   &&\frac{-\lambda
L^2}{128\pi^2}
    \Bigg[f(a)-f(b)+2f(\frac{L-a}{2})-2f(\frac{L-b}{2})\nonumber\\
   &&+\frac{2m^2}{\pi}\left((B(a)-B(b)
   +2B(\frac{L-a}{2})-2B(\frac{L-b}{2})\right)
   \int_{0}^{\infty}ds\sqrt{1+s^{2}}\Bigg]
\end{eqnarray}
It is important to note that all these cancelations are easily
accomplished using our supplementary box renormalization scheme. On
a minor note, it is interesting to note that the changes of the
variables leading to the cancelation of infinities are,
surprisingly, equivalent to setting different cutoff regularizations
on the upper limits of the integrals. Equation (\ref{Res.d=3..}) is
our main equation for the NLO Casimir energy, and appears in the
text as eq.~(\ref{e27:corr.cas.E}), and is analyzed further there.

\end{document}